\documentclass[final]{aipproc}
\layoutstyle{6x9}
\usepackage{epsfig}
\usepackage{amssymb,amsfonts}
\begin{document}
\title{Parity doublers in chiral potential quark models}

\author{Yu. S. Kalashnikova}{address={Institute of Theoretical and Experimental Physics, 117218,
B.Cheremushkinskaya 25, Moscow, Russia}}

\author{A. V. Nefediev}{address={Institute of Theoretical and Experimental Physics, 117218,
B.Cheremushkinskaya 25, Moscow, Russia}}

\author{J. E. F. T. Ribeiro}{address={Centro de F\'\i sica das Interac\c c\~oes Fundamentais
(CFIF), Departamento de F\'\i sica, Instituto Superior T\'ecnico, Av.
Rovisco Pais, P-1049-001 Lisboa, Portugal}}

\newcommand{\low}[1]{\raisebox{-1mm}{$#1$}}
\newcommand{\up}[1]{\raisebox{+2mm}{$#1$}}
\newcommand{\vp}{\varphi}
\newcommand{\ds}{\displaystyle}
\newcommand{\vx}{{\vec x}}
\newcommand{\vy}{{\vec y}}
\newcommand{\vz}{{\vec z}}
\newcommand{\vk}{{\vec k}}
\newcommand{\vq}{{\vec q}}
\newcommand{\vpp}{{\vec p}}
\newcommand{\vn}{{\vec n}}
\newcommand{\vg}{{\vec \gamma}}
\newcommand{\be}{\begin{equation}}
\newcommand{\bea}{\begin{eqnarray}}
\newcommand{\ee}{\end{equation}}
\newcommand{\eea}{\end{eqnarray}}
\newcommand{\loww}[1]{\raisebox{-1.5mm}{$#1$}}
\newcommand{\lmn}{\mathop{\sim}\limits_{n\gg 1}}
\newcommand{\vpint}{\int\makebox[0mm][r]{\bf --\hspace*{0.13cm}}}
\newcommand{\too}{\mathop{\to}\limits_{N_C\to\infty}}


\begin{abstract}
The effect of spontaneous breaking of chiral symmetry over the spectrum of
highly excited hadrons is addressed in the framework of a microscopic
chiral potential quark model (Generalised Nambu-Jona-Lasinio model) with a
vectorial instantaneous quark kernel of a generic form. A heavy-light
quark-antiquark bound system is considered, as an example, and the Lorentz
nature of the effective light-quark potential is identified to be a pure
Lorentz-scalar, for low-lying states in the spectrum, and to become a pure
spatial Lorentz vector, for highly excited states. Consequently, the
splitting between the partners in chiral doublets is demonstrated to
decrease fast in the upper part of the spectrum so that neighboring states
of an opposite parity become almost degenerate. A detailed microscopic
picture of such a ``chiral symmetry restoration" in the spectrum of highly
excited hadrons is drawn and the corresponding scale of restoration is
estimated.
\end{abstract}

\maketitle

Chiral symmetry is known to be broken spontaneously in QCD, and this phenomenon plays an important role for low--lying hadrons. The (almost) massless chiral pion constitutes the most prominent example of chiral symmetry breaking manifestation in the hadronic spectrum. In the meantime, considering quite general quantum--mechanical principles \cite{Glozman1}, one can argue that this manifestation should asymptotically disappear for highly excited states \cite{Glozman}. Although this property was studied before using various effective and phenomenological approaches \cite{Bardeen,Nowak,Swanson}, the microscopic picture of such an effective chiral symmetry restoration in highly excited hadrons was not disclosed so far. In the meantime, a model exists which meets all the requirements necessary to reproduce this phenomenon \cite{parity2,parity3}. This is the Generalised Nambu-Jona-Lasinio (GNJL) model for QCD \cite{Orsay,Orsay2,Lisbon}. Indeed, the model is chirally symmetric (in the chiral limit) and is able to describe microscopically the phenomenon of spontaneous breaking of chiral symmetry in the vacuum. It is intrinsically relativistic and contains confinement, so it should be able to address the problem of highly excited hadrons. The model is described by the Hamiltonian:
\be 
H=\int
d^3 x q^\dagger(x)(-i\vec{\alpha}\vec{\bigtriangledown})q(x)+ \int
d^3xd^3y \left[q^\dagger(x)\frac{\lambda^a}{2}q(x)\right]
V(\vec{x}-\vec{y})
\left[q^\dagger(y)\frac{\lambda^a}{2}q(y)\right],
\label{H}
\ee 
where, for the sake of simplicity, we stick to the simplest form of the Hamiltonian (\ref{H}) compatible with the requiremets of confinement and chiral symmetry breaking. The standard approach used in this kind of models is the Bogoliubov--Valatin transformation from bare to dressed quarks parametrised with the help of the chiral angle $\vp_p$ \cite{Lisbon}.
The mass--gap equation which defines the profile of the chiral angle appears from the requirement that the quadratic part of the normally ordered Hamiltonian (\ref{H}) is diagonal, $:H_2: \propto b^\dagger b+d^\dagger d$. Then two phases of the theory can be identified: the unbroken phase with $\vp_p\equiv 0$, and the broken phase, with the chiral angle taking the form depicted in Fig.~1. The broken phase possesses a lower vacuum energy and is therefore the true vacuum state of the theory.

\begin{figure}[t]
\begin{tabular}{ccc}
\epsfig{file=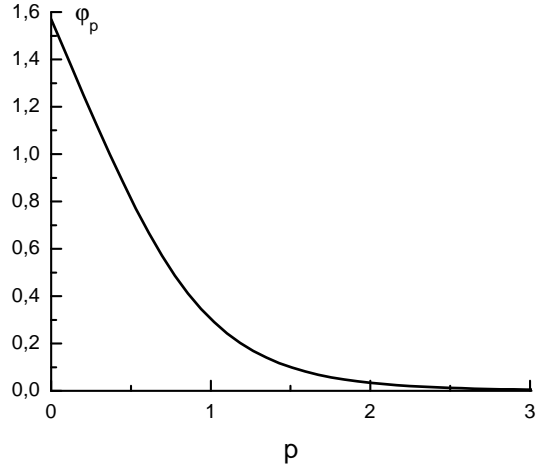,width=7cm}
\caption{The typical profile of the chiral angle --- solution to the mass--gap equation}
\end{tabular}
\end{figure}

Every mesonic state in the model is described by a pair of wave functions: one responsible for the time--forward and the other --- for the time--backward motion of the quark--antiquark pair in the meson \cite{Orsay,Orsay2,Lisbon,nr}. To make things simpler, we consider one quark infinitely heavy, so that the heavy--light meson w.f. becomes one--component and it obeys the Schr{\" o}dinger-like equation, in momentum space, \cite{parity2}
\be
E_p\psi(\vec{p})+\int\frac{d^3k}{(2\pi)^3}V(\vpp-\vk)\left[C_pC_k+({\vec \sigma}\hat{\vpp})({\vec \sigma}\hat{\vk})S_pS_k\right]\psi(\vec{k})=E\psi(\vec{p}),
\label{se}
\ee
where $E_p$ is the dressed--quark dispersive law; $C_p=\cos\frac12\left(\frac{\pi}{2}-\vp_p\right)$, 
$S_p=\sin\frac12\left(\frac{\pi}{2}-\vp_p\right)$. The effect of chiral symmetry restoration in highly excited hadrons can be quite naturally exemplified by the behavious of the splitting between the opposite--parity states $\psi(\vec{p})$ and $\psi'(\vec{p})$ as the excitation number increases (in this work, we keep the radial excitation number fixed and increase the light--quark angular momentum). It is easy to notice that, for highly excited mesons, the mean interquark mometum is large, so that $C_p\approx S_p\approx\frac{1}{\sqrt{2}}$ and the resulting bound--state equation becomes symmetric with respect to the substitution $\psi(\vec{p})\to\psi'(\vec{p})=({\vec \sigma}\hat{\vpp})\psi(\vec{p})$. Thus these opposite--parity states become degenerate. It is easy to trace the origin of such a degeneracy. Indeed, the splitting between $\psi(\vec{p})$ and $\psi'(\vec{p})$ comes from the difference 
$C_p^2-S_p^2=\sin\vp_p$, so that this is the large--momentum behaviour of the chiral angle to play a decisive role for the symmetry restoration --- see Fig.~1.
It is instructive to approach the same problem from the point of view of the Lorentz nature of the effective interquark interaction. Thus we perform Foldy counter--rotation of Eq.~(\ref{se}) to arrive at the Dirac--like equation for the light quark:
\be
({\vec \alpha}\vpp+\beta m)\Psi(\vx)+\frac12\int d^3z \;U(\vx-\vz)[V(\vx)+V(\vz)-V(\vx-\vz)]\Psi(\vz)=E\Psi(\vx).
\label{de}
\ee
The Lorentz nature of confinement in this equation follows from the matrix structure of $U$ which takes the form, in momentum space,
\be
U(\vec{p})=\beta\sin\vp_p+({\vec \alpha}\hat{\vpp})\cos\vp_p.
\ee
We conclude therefore that, for low--lying states, the effective scalar interaction dominates. It is chirally nonsymmetric and thus chiral symmetry breaking manifests itself in this part of the spectrum (this regime was studied in detail in \cite{Simonov}). On the contrary, for highly excited states, the effective interquark interaction (asymptotically) becomes purely vectorial, so that chiral symmetry is an approximate symmetry of the interaction and thus approximate chiral multiplets appear in the spectrum of excited states. Notice that the transition between the two regimes is governed by $\sin\vp_p$ which is known to be exactly the quantity ``responsible"' for chiral symmetry breaking in this class of models.

Below we exemplify our qualitative conclusions drawn above with exact calculations performed for the harmonic oscillator potential case, $V(r)=K_0^3r^2$, with $K_0$ being the only dimensional parameter of the model \cite{Orsay,Orsay2,Lisbon}. For this potential, the mass--gap equation and the bound--state euqation become ordinary second--order differential equations. As soon as the mass--gap equation is solved, we fix the value of the scale $K_0$ by evaluating the chiral condensate, $\langle\bar{q}q\rangle=-\frac{3}{\pi^2}\int_0^\infty dp\;p^2\sin\vp_p\approx-(0.51K_0)^3$, and setting it equal to the standard value of $-(250MeV)^3$ (for future references we call this scale the BCS scale, $\Lambda_{\rm BCS}=250MeV$). This gives $K_0=490MeV$.
The bound--state Eq.~(\ref{se}) can be written now, in momentum space, as
\be
-K_0^3u''+V(p)u=Eu,\quad\psi(\vpp)=\Omega_{jlm}(\hat{\vec{p}})\frac{u(p)}{p},
\label{se2}
\ee
with the effective potential
\be
V(p)=E_p+K_0^3\left[\frac14\vp_p^{\prime 2}+\frac{(j+1/2)^2}{p^2}+\frac{\kappa}{p^2}\sin\vp_p\right],\quad\kappa=\pm\left(j+1/2\right).
\ee
This allows us to extract the difference between the potentials operative for the opposite--parity states:
\be
\Delta V=-\frac{2(j+1/2)K_0^3}{p^2}\sin\vp_p,
\ee
and which is responsible for the splitting between such states. As it was anticipated before, this potential is directly proportional to $\sin\vp_p$ which vanishes as $p\to\infty$, and so does the splitting \cite{parity2}. It is instructive to compare this situation with the Salpeter equation
$[\sqrt{p^2+m^2}+K_0^3r^2]\psi(\vec{x})=E\psi(\vec{x})$ which leads to the splitting potential $\Delta V=-\frac{2(l+1)K_0^3}{p^2}$. Results of numerical calculations are plotted in the form of Regge trajectories, in Fig.~2 (left plot) for the radial quantum number $n=0$. It is clearly seen from this plot that, for the full bound--state equation the trajectories corresponding to the parity doublers merge thus reflecting the effect of chiral symmetry restoration.

\begin{figure}[t]
\begin{tabular}{ccc}
\epsfig{file=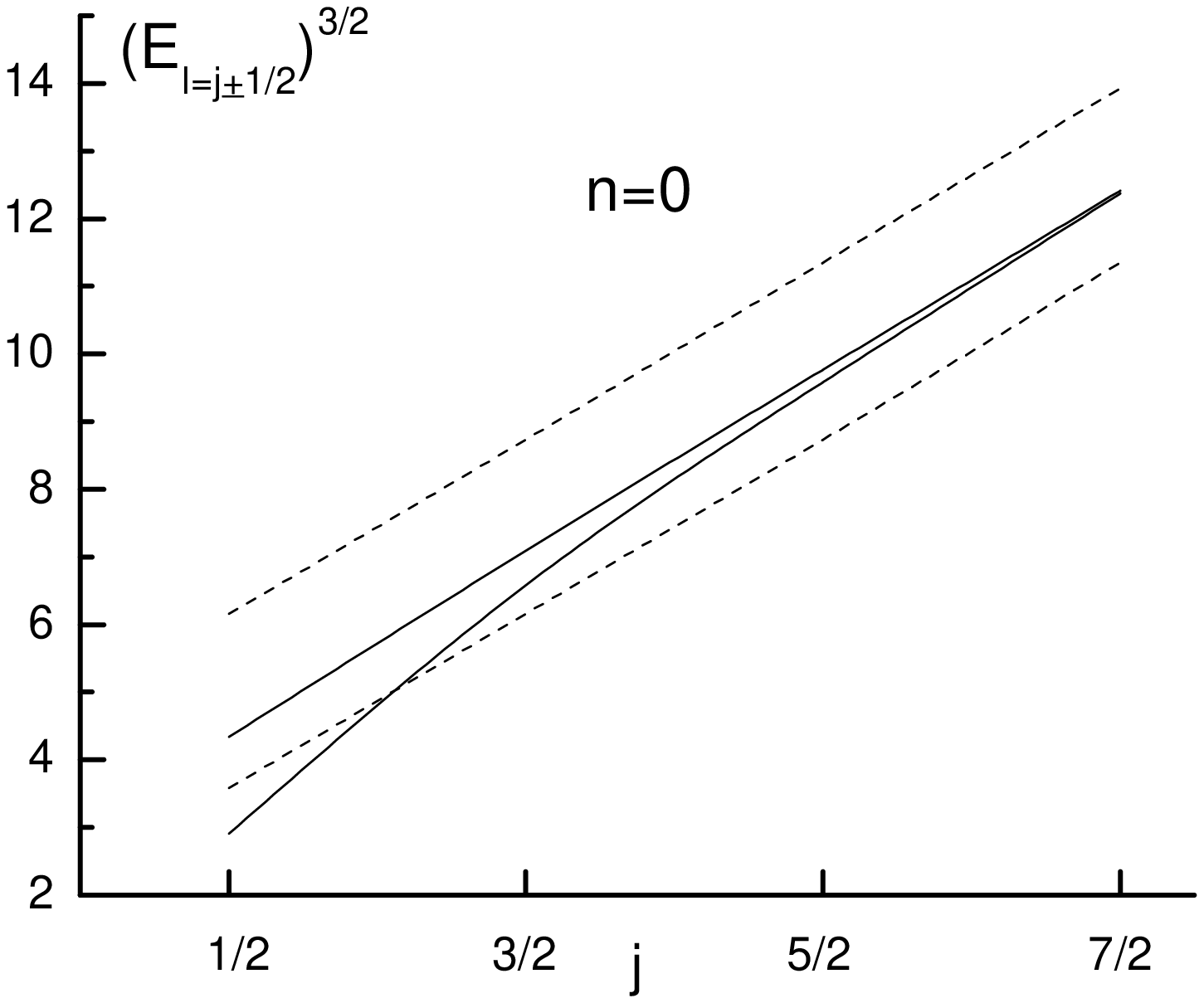,width=6.5cm,width=6.5cm}&&\epsfig{file=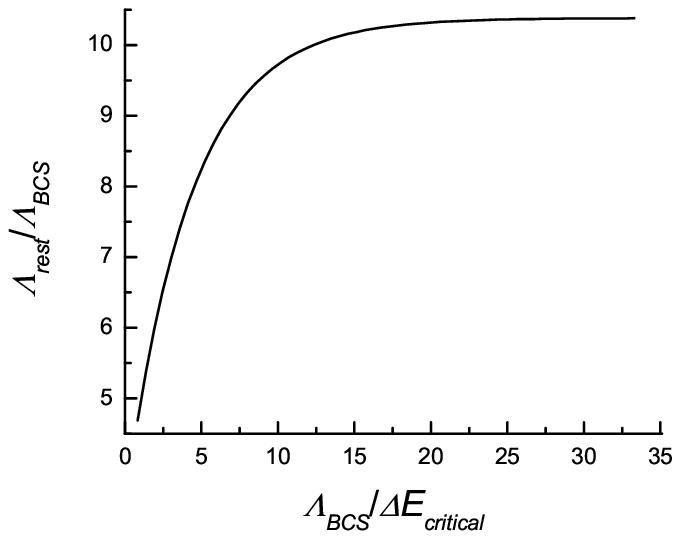,width=6.5cm}
\end{tabular}
\caption{Left plot: Regge trajectories for the full bound--state Eq.~(\ref{se2}) (solid line) and for the Salpeter equation (dashed line). Right plot: the restoration scale against the splitting.}
\end{figure}

Finally, we estimate the restoration scale \cite{parity2}. To this end we notice that the splitting in chiral doublets appears of order of the BCS scale in the lowest part of the spectrum and, above some scale --- the restoration scale $\Lambda_{\rm rest}$, it is much smaller than $\Lambda_{\rm BCS}$. We use this as the definition of $\Lambda_{\rm rest}$. Then, from the right plot at Fig.~2, one can conclude that $\Lambda_{\rm rest}\approx 10 \Lambda_{\rm BCS}\approx 2.5GeV$.

Therefore we observe that the Generalised Nambu--Jona-Lasinio model provides a clear microscopical pattern of chiral symmetry restoration in the spectrum of highly excited hadrons. This effective asymtotic restoration happens rather fast and the model is able to predict a reasonable value of the restoration scale in agreement with other estimates found in the literature \cite{Swanson}.
\bigskip

The research of A.N. and Yu.K. was supported by the grants RFFI-05-02-04012-NNIOa, DFG-436 RUS 113/820/0-1(R), NSh-843.2006.2,
by the Federal Programme of the Russian Ministry of Industry, Science, and
Technology No. 40.052.1.1.1112, and by the Russian Governmental Agreement
N 02.434.11.7091.


\begin{thebibliography}{99}
\expandafter\ifx\csname natexlab\endcsname\relax\def\natexlab#1{#1}\fi
\providecommand{\enquote}[1]{``#1''}
\expandafter\ifx\csname url\endcsname\relax
  \def\url#1{\texttt{#1}}\fi
\expandafter\ifx\csname urlprefix\endcsname\relax\def\urlprefix{URL }\fi

\bibitem{Glozman1} L. Ya. Glozman, {\em Int. J. Mod. Phys.} {\bf A21}, 475 (2006) 475.
\bibitem{Glozman} L. Ya. Glozman, {\em Phys. Lett.} {\bf B475}, 329 (2000);
T. D. Cohen and L. Ya. Glozman, {\em Int. J. Mod. Phys.} {\bf A17}, 1327 (2002); {\em Phys. Rev.} {\bf D65}, 016006 (2002).
\bibitem{Bardeen} W. A. Bardeen, E. J. Eichten, and C. T. Hill, {\em Phys. Rev.} {\bf D68},
054024 (2003).
\bibitem{Nowak} M. Nowak, M. Rho, and I. Zahed,  {\em Acta Phys. Polon.} {\bf B35},
2377 (2004).
\bibitem{Swanson} E. S. Swanson, {\em Phys. Lett.} {\bf B582}, 167 (2004).
\bibitem{parity2} Yu. S. Kalashnikova, A. V. Nefediev, and J. E. F. T. Ribeiro, {\em Phys. Rev.} {\bf D72}, 034020 (2005).
\bibitem{parity3} L. Ya. Glozman, A. V. Nefediev, and J. E. F. T. Ribeiro, {\em Phys. Rev.} {\bf D72}, 094002 (2005).
\bibitem{Orsay} A. Amer, A. Le Yaouanc, L. Oliver, O. Pene, and J.-C.
Raynal, {\em Phys. Rev. Lett.} {\bf 50}, 87 (1983); A. Le Yaouanc, L. Oliver,
O. Pene, and J.-C. Raynal, Phys. Lett. B {\bf 134}, 249 (1984); {\em Phys.
Rev.} {\bf D29}, 1233 (1984).
\bibitem{Orsay2} A. Le Yaouanc, L. Oliver, S. Ono, O. Pene and
J.-C. Raynal, {\em Phys. Rev.} {\bf D31}, 137 (1985).
\bibitem{Lisbon} P. Bicudo and J. E. Ribeiro, {\em Phys. Rev.} {\bf D42},
1611 (1990); {\it ibid.}, 1625 (1990); {\it ibid.}, 1635 (1990).
\bibitem{nr} A. V. Nefediev and J. E. F. T. Ribeiro, {\em Phys. Rev.} {\bf D70}, 094020 (2004).
\bibitem{Simonov} Yu. A. Simonov, {\em Yad. Fiz.} {\bf 60}, 2252 (1997) [{\em Phys. Atom. Nucl.} {\bf 60}, 2069 (1997)].
\end{thebibliography}
\end{document}